\documentclass[twocolumn,showpacs,preprintnumbers,amsmath,amssymb,prb]{revtex4}

\usepackage{graphicx}
\usepackage{dcolumn}
\usepackage{bm}
\def\Vec#1{\mbox{\boldmath $#1$}}
\def\kv{\mbox{\boldmath $ k$}}

\def\pv{\mbox{\boldmath $ p$}}

\def\rv{\mbox{\boldmath $r$}}

\def\up{\uparrow}
\def\down{\downarrow }

\begin{document}


\title{Positions of Point-Nodes in Borocarbide Superconductor YNi$_2$B$_2$C}

\author{Yuki Nagai$^{1,2}$}

\author{Yusuke Kato$^{1,3}$}%
\author{Nobuhiko Hayashi$^{4,5}$}
\author{Kunihiko Yamauchi$^6$}
\author{Hisatomo Harima$^7$}
\affiliation{%
$^{1}$Department of Physics, University of Tokyo, Tokyo 113-0033, Japan\\
$^{2}$Condensed Matter Theory Laboratory, RIKEN, Saitama, 351-0198, Japan\\
$^{3}$Department of Basic Science, University of Tokyo, Tokyo 153-8902, Japan\\
$^{4}$CCSE, Japan  Atomic Energy Agency, 6-9-3 Higashi-Ueno, Tokyo 110-0015, Japan\\
$^{5}$CREST JST, 4-1-8 Honcho, Kawaguchi, Saitama 332-0012, Japan\\
$^{6}$CNR-INFM, CASTI Regional Lab, I-67010 Coppito (L'Aqulia), Italy\\
$^{7}$Department of Physics, Kobe University, Nada, Kobe 657-8501, Japan}%

\date{\today}

\begin{abstract}
To determine the superconducting gap function of YNi$_2$B$_2$C, 
we calculate the local density of states (LDOS) around a single vortex core with
 the use of 
 Eilenberger theory and the band structure calculated by local density approximation
assuming 
various gap structures with point-nodes at different positions.
We also calculate the angular-dependent heat capacity in the vortex state on the basis of the Doppler-Shift method.
Comparing our results with the STM/STS experiment, the angular-dependent heat capacity and thermal conductivity,
we propose the gap-structure of YNi$_2$B$_2$C, which has the point-nodes and gap minima along $\langle 110 \rangle$. 
Our gap-structure is consistent with all results of angular-resolved experiments.
\end{abstract}

\maketitle
\section{Introduction}
The discovery of the nonmagnetic borocarbide superconductor YNi$_2$B$_2$C\cite{Cava} has considerable attention 
because of the growing evidence for highly anisotropic superconducting gap 
and high superconducting transition temperature $15.5$ K. 
The boron isotope effect supports the classification of this material as electron-phonon mediated superconductor.\cite{Lawrie,Cheon}
At an early stage, from specific heat, thermal conductivity, Raman scattering and photoemission spectroscopy 
experiments on YNi$_2$B$_2$C, a highly anisotropic gap function was concluded.\cite{Muller}
In recent years, Maki {\it et al.}\cite{MakiThal} theoretically suggested 
that the gap symmetry of this material is $s$+$g$ wave and the gap function has zero points (point-nodes) in momentum space.
Motivated by this prediction, 
field-angle-dependent heat capacity (FAD heat capacity)\cite{Park} and angular variation
 of the thermal transport (AV thermal transport)\cite{Izawa}
on YNi$_2$B$_2$C have been measured and these results were considered to be consistent with this prediction.
The FAD heat capacity and the AV thermal conductivity in $\Vec{H}$ 
rotated within the $ab$ plane show a fourfold oscillation with narrow cusps 
because of the presence of nodal quasiparticles subject to Doppler shifts.
These results suggest that the gap function has point-nodes along the $a$ and $b$ axes.
However, in the analysis of both experiments, the isotropic Fermi surface (FS)  was assumed.
YNi$_2$B$_2$C has highly anisotropic Fermi surfaces (FSs).\cite{Lee,Singh,Yamauchi}

\begin{figure}
\includegraphics[width = 6cm]{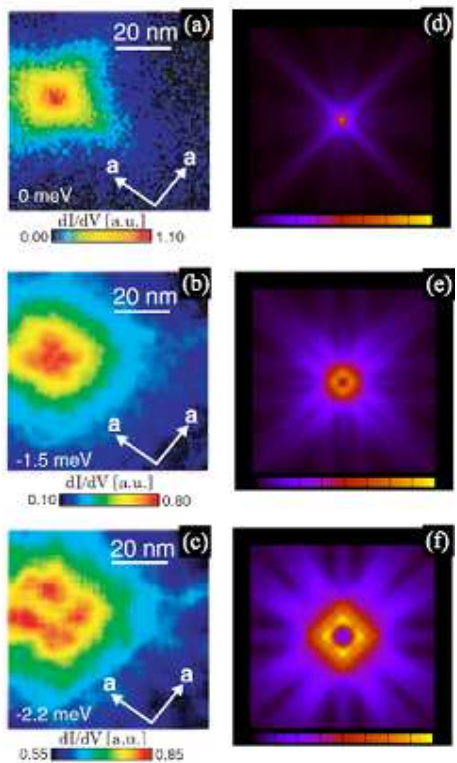}
\caption{\label{fig:LDOSs}(Color Online) Tunneling conductance images observed by Nishimori {\it et al.} at 0.46K 
in 0.07 T ($72 \times 72$nm) for the bias voltage (a) 0meV, (b)-1.5meV, (c)-2.2meV. 
In this figure, 
we quoted from Fig.5 (a), (c) and (e) in ^^ {\it First Observation of the Fourfold-symmetric and Quantum Regime Vortex Core 
in YNi$_2$B$_2$C by Scanning Tunneling Microscopy and Spectroscopy}' 
(H. Nishimori {\it et al}, Journal of the Physical Society of Japan {\bf 73}, 3247 (2004)). 
The LDOS images calculated for (d)$\epsilon/\Delta_{\infty} = 0$, (e)0.15, (f)0.3, where $4\sqrt{2} \xi_0 \times 4\sqrt{2} \xi_0$ is shown.
Here, $\Delta_{\infty}$ denotes a pair-potential in the bulk region averaged on the Fermi surface and $\xi_0$ denotes a coherent length.
The smearing factor is $\delta = 0.05 \Delta_{\infty}$}
\end{figure}

The local density of states (LDOS) in the isolated vortex of YNi$_2$B$_2$C was measured at 0.46K 
by the scanning tunneling microscopy and spectroscopy (STM/STS).\cite{Nishimori}
The vortex core was found to be fourfold star-shaped in real space (see Fig.~\ref{fig:LDOSs}(a)-(c)).
Near $E = 0$meV, the LDOS extends toward $\langle 100 \rangle$ (the $a$-axis) and has a peak at the center of the vortex core (Fig.~\ref{fig:LDOSs}(a)).
With increasing energy, the peak of the LDOS splits into four peaks toward $\langle 110 \rangle$ (Fig.~\ref{fig:LDOSs}(b),(c)).
The anisotropic spatial pattern of the LDOS around a vortex core is a consequence of the anisotropic pair-potential and FS.
Many theoretical studies have been done since the LDOS of NbSe$_2$ can be observed by STM/STS.\cite{Gygi, Hayashi, Schopohl,Ichioka,Caroli,Kramer,Nagai_ce,Nagai,Hess}
We have calculated the LDOS pattern around a 
vortex core of YNi$_2$B$_2$C, assuming that the gap symmetry is $s$+$g$-wave and the FS is isotropic.\cite{Nagai}
Since this result explains the STM results only partly, 
obviously the calculations of the LDOS with a highly anisotropic FS are needed.

The band structure for YNi$_2$B$_2$C has been calculated, on the basis of the local density approximation (LDA).
Lee {\it et al.}\cite{Lee} and Singh \cite{Singh} obtained three bands (17th, 18th and 19th) with mainly Ni-3d and Y-4d 
character which cut across the Fermi level 
by using the linearized muffin-tin orbital method and the general potential linearized augmented plane wave method, respectively.
In recent years, band structure calculations were carried out by Yamauchi {\it et al.},\cite{Yamauchi} 
who are some of the authors,  by 
using a full potential LAPW(FLAPW) method. 
They have carried out LDA-based calculations with a certain modification
to describe well the experimental FSs which are obtained by de Haas 
van Alphen effect.
In this modification, Y-d and Ni-d levels are shifted upward from
the original LDA levels by 0.11Ry and 0.05 Ry, respectively.
Such the modification have been successful in the FS calculations in 
LaB$_6$,\cite{Harima1} YbAl$_3$\cite{Ebihara} and LaRh$_3$B$_2$.\cite{Harima2}
This shift may originate mainly in the self-interaction and/or the non-local corrections to the LDA.
They have shown that the density of states from 17th band,
which is related to a multiple connected electron FS, has a sharp  
peak at Fermi energy.
This peak comes from a van Hove singularity around (1/5, 1/5, 0)  
point in k-space as we will explain later in this paper.
They have suggested that this singularity may lead large electron- 
phonon coupling locally, and give rise to anisotropic gap behavior in  
the superconducting state.

The purpose of this paper is to determine the gap structure consistent with 
the STM/STS, 
the FAD heat capacity and the AV thermal transport experiments.
We show the results of four calculations considering the band structure calculations.
First, we calculate the LDOS pattern around a single vortex 
on the basis of the quasiclassical Eilenberger theory.\cite{Nagai}
We assume various gap structures, 
and finally obtain the gap structure which can explain the STM experiment.
Second, we calculate the density of states in weak magnetic fields 
in the $a$-$b$ plane on the basis of the Doppler-shift method.\cite{Volovik}
We show that 
the FAD heat capacity and the AV thermal transport experiments are consistent with 
the angular dependence of the density of states from the gap structure consistent with the STM experiment 
in the presence of magnetic fields.
Third, we show that the density of states 
from this gap structure is consistent with the STM experiment 
in the absence of magnetic fields.
As results, we show the most suitable gap structure and positions of point-nodes.
Finally, we show the LDOS patterns in the magnetic fields which tilts from the crystal $c$ axis. 
With angular dependence of the LDOS, we can obtain more rich information about the gap structure.
\section{Quasiclassical Theory of Superconductivity}
We consider the pair-potential:\cite{Ueda} 
\begin{equation}
\Delta_{\up \down}(\Vec{R},\kv) = \int d\Vec{q} \Delta_{\up \down}(\Vec{q},\kv) 
	e^{{\rm i} \Vec{q} \cdot \Vec{R}},
\end{equation}
\begin{equation}
\Delta_{\up \down}(\Vec{q},\kv) = - \sum_{\kv',s_1,s_2} V_{\up,\down,s_1,s_2}(\kv,\kv') 
	\langle 
        	a_{\frac{\Vec{q}}{2}+\kv',s_1} a_{\frac{\Vec{q}}{2}-\kv',s_2}
                \rangle, 
\end{equation}
where $\kv$ corresponds to the internal degree of freedom of the pairing state. 
Here, $\Vec{R}$ is the center-of-mass coordinate of Cooper pairs, 
$V_{\up,\down,s_1,s_2}(\kv,\kv')$ is the interaction between two electrons, 
$a_{\Vec{q}/2\pm \kv',s}$ is the annihilation operator for the quasiparticle states with spin $s$ and 
momentum $\Vec{q}/2 \pm \kv'$ 
and we use units in which $\hbar = k_{\rm B} = 1$.
We assume the weak coupling interaction so that the pair-potential is not zero only near the Fermi surface.
We consider pair-potential written as $\hat{\Delta}(\Vec{R},\kv) = {\rm i} \psi(\kv) \hat{\sigma}_y A(\Vec{R})$ for singlet pairing.
Here $A(\Vec{R})$ is a function of $\Vec{R}$. 
If the coherence length at the zero temperature $\xi_0$ is large compared to the Fermi wave 
length ($\sim 1/ k_{\rm F}$), 
we can calculate the LDOS around a vortex core on the basis of the quasiclassical theory of superconductivity.\cite{Eilen,Serene,Larkin}
We consider the quasiclassical Green function $\check{g}$ that has matrix elements in the Nambu (particle-hole) space as 
\begin{eqnarray}
\check{g}(\rv,\kv,{\rm i}\omega_n) = \left(
\begin{array}{cc}
g & f \\
-\bar{f} & \bar{g}
\end{array}
\right),
\end{eqnarray}
where $\omega_n$ is the Matsubara frequency.
The equation of motion for $\check{g}$ called Eilenberger equation for singlet pairing is written as
\begin{eqnarray}
{\rm i} \Vec{v}_{\rm F}(\kv) \cdot \nabla \check{g} + \left[
\left(
\begin{array}{cc}
 {\rm i} \omega_n& -\Delta \\
\Delta^{\ast} & -{\rm i} \omega_n
\end{array}
\right)
, \check{g}
 \right]
  =0.
\end{eqnarray}
Here, $\Vec{v}_{\rm F}$ is the Fermi velocity 
and $[\check{a},\check{b}]$ denotes the commutator $\check{a} \check{b}- \check{b} \check{a}$.
The Green function $\check{g}$ satisfies the normalization condition: $\check{g}^2 = \check{1}$, 
where $\check{1}$ is a $2 \times 2$ unit matrix.
Considering clean superconductors in the type II limit,\cite{Muller} we neglect the self-energy part of Green function and the vector potential.

The local density of states with the isotropic Fermi surface is given by 
\begin{equation}
\nu(\rv,\epsilon) = - \nu(0) \int \frac{d \Omega_k}{ 4 \pi} {\rm Re} \; {\rm tr} (\hat{g}^{\rm R}). \label{eq:ldos}
\end{equation}
Here, $\hat{g}^{\rm R}$ is the retarded Green function $\hat{g}^{\rm R} = \hat{g}({\rm i} \omega_n \rightarrow \epsilon+{\rm i} \delta)$
 and $\nu(0)$ denotes the density of states on Fermi surface 
in the normal metallic state.
The local density of states with the anisotropic Fermi surface is given by 
\begin{equation}
\nu(\rv,\epsilon) = - \int \frac{d S_{\rm F} }{2 \pi^2 v_{\rm F}} 
 {\rm Re} \: {\rm tr} (\hat{g}^{\rm R}). \label{eq:ldosa}
\end{equation}
Here, $d S_{\rm F}$ is the Fermi-surface area element and 
$v_{\rm F}$ is the modulus of Fermi velocity.
\section{Riccati Formalism and Kramer-Pesch Approximation}

The Eilenberger equation can be simplified by introducing a parametrization for the propagators that 
satisfy the normalization condition. 
Propagators are defined as $\check{P}_{\pm} = (\check{1} \mp \check{g})/2$, which were 
originally introduced in the studies of vortex dynamics.\cite{Shelankov,eschrig} 
Using these propagators, we obtain the scalar equations expressed in Riccati formalism as follows: 
\begin{eqnarray}
\Vec{v}_{\rm F} \cdot \Vec{\nabla} a_+ + 2 \omega_n a_+ + a_+ \Delta^{\ast} a_+ - \Delta &=& 0, \label{eq:ar}\\
\Vec{v}_{\rm F} \cdot \Vec{\nabla} b_- - 2 \omega_n b_- - b_- \Delta b_- + \Delta^{\ast} &=& 0, \label{eq:br}
\end{eqnarray}
where
\begin{eqnarray}
\check{g} &=& - \check{N} \left(
\begin{array}{cc}
(1 -a_+ b_-) & 2 {\rm i} a_+ \\
- 2 {\rm i} b_- & -(1 - b_- a_+)
\end{array}
\right), \\
\check{N} &=& \left(
\begin{array}{cc}
(1+a_+ b_-)^{-1} & 0 \\
0 & (1 + b_- a_+)^{-1}
\end{array}
\right).
\end{eqnarray}
Since these equations (\ref{eq:ar}) and (\ref{eq:br}) contain $\Vec{\nabla}$ only through $\Vec{v}_{\rm F} \cdot \Vec{\nabla}$, 
they reduce to a one-dimensional problem on a straight line, 
the direction of which is given by that of the Fermi velocity $\Vec{v}_{\rm F}$.
We consider a single vortex along the $Z$ axis parallel to the crystal $c$ axis. 
When we take the $X$ axis on the $a$-$b$ plane, the $Y$ axis is determined automatically. 
We denote by $\hat{\Vec{a}}_{\rm M}$, $\hat{\Vec{b}}_{\rm M}$ and $\hat{\Vec{c}}_{\rm M}$ the unit vectors along $X$, $Y$, $Z$ axis, 
respectively.
We introduce the vector written as 
\begin{equation}
\Vec{R} = X \hat{\Vec{a}}_{\rm M} + Y \hat{\Vec{b}}_{\rm M} + Z \hat{\Vec{c}}_{\rm M}.
\end{equation}
The origin $\Vec{R} = 0$ is put on the vortex center.
Because of a translational symmetry along the $Z$ axis, 
the pair-potentials $\Delta(\Vec{R},\kv)$, $\Delta^{\ast}(\Vec{R},\kv)$ do not depend on $Z$ in the Riccati equations (\ref{eq:ar}) and (\ref{eq:br}),
and hence $a_+$ and $b_-$ depend on $\Vec{R}$ only through $X$ and $Y$.
As a result, $\Vec{v}_{\rm F} \cdot \Vec{\nabla}$ in the Riccati equations can be replaced by 
\begin{equation}
\Vec{v}_{\rm F} \cdot \Vec{\nabla} \rightarrow  \Vec{v}_{\rm F \perp} \cdot \Vec{\nabla} =  v_X \frac{\partial}{\partial X} + v_Y \frac{\partial}{\partial Y},
\end{equation}
with $v_X = \Vec{v}_{\rm F} \cdot \hat{\Vec{a}}_{\rm M}$ and  $v_Y = \Vec{v}_{\rm F} \cdot \hat{\Vec{b}}_{\rm M}$.
Here, $\Vec{v}_{\rm F \perp}$ is the vector perpendicular to the $Z$ axis by projecting the Fermi velocity $\Vec{v}_{\rm F}$
on the $\hat{\Vec{a}}_{\rm M}$-$\hat{\Vec{b}}_{\rm M}$ plane.
We introduce $x$, $y$, $r$ by 
\begin{eqnarray}
\Vec{R} &=& x \hat{\Vec{v}} + y \hat{\Vec{u}} + Z \hat{\Vec{c}}_{\rm M}, \\
\Vec{r}  &\equiv& x \hat{\Vec{v}} + y \hat{\Vec{u}} , \\
r &\equiv& \sqrt{x^2+y^2} = \sqrt{X^2+Y^2},
\end{eqnarray}
with 
\begin{eqnarray}
\left(
\begin{array}{c}
\hat{\Vec{v}}  \\
 \hat{\Vec{u}}
\end{array}
\right)
 &\equiv& \left(
\begin{array}{cc}
\cos \theta_v & \sin \theta_v \\
- \sin \theta_v & \cos \theta_v
\end{array}
\right)\left(
\begin{array}{c}
\hat{\Vec{a}}_{\rm M}  \\
 \hat{\Vec{b}}_{\rm M}
\end{array}
\right).
\end{eqnarray}
Here, $\theta_v$ is the angle between $\hat{\Vec{a}}_{\rm M}$  and the velocity $\Vec{v}_{\rm F \perp}$.
The resultant Riccati equations are then written as 
%
\begin{eqnarray}
v_{\rm F \perp} h(\Vec{k}_{\rm F}) \frac{\partial a_+}{\partial x} + 2 \omega_n a_+ + a_+^2 \Delta^{\ast} - \Delta &=& 0, \label{eq:ari}\\
v_{\rm F \perp} h(\Vec{k}_{\rm F}) \frac{\partial b_-}{\partial x} - 2 \omega_n b_- - b_-^2 \Delta  + \Delta^{\ast} &=& 0 \label{eq:bri}.
\end{eqnarray}
Here, $v_{\rm F \perp}$ and $h(\Vec{k}_{\rm F})$ denote by 
\begin{eqnarray}
v_{\rm F \perp} &\equiv& \frac{\int dS_{\rm F} |\Vec{v}_{\rm F \perp}|}{ \int d S_{\rm F}} \\
v_{\rm F \perp} h(\kv_{\rm F}) &\equiv& \sqrt{v_X^2(\kv_{\rm F})+ v_Y^2(\kv_{\rm F})}. 
\end{eqnarray}

An approximate expression for the quasiclassical Green function near a vortex core for low energy was 
analytically obtained by Kramer and Pesch.\cite{Kramer} 
We call their approximation ^^ Kramer-Pesch approximation(KPA)' in present paper. 
Eschrig has also obtained the quasiclassical Green function near a vortex core 
in the first order of impact parameter and energy 
with use of the Riccati formalism.\cite{eschrig} 
The method by Eschrig is equivalent to KPA.
Using KPA,  
we can calculate the quasiclassical Green function 
around the vortex core in a low energy region ($|\omega_n| \ll |\Delta_{\infty}|$).
Here, $\Delta_{\infty}$ denotes a pair potential in the bulk region
\begin{equation}
\Delta_{\infty} \equiv \frac{\int dS_{\rm F} \lim_{\Vec{r} \rightarrow \infty}\Delta(\Vec{r},\kv)}{\int dS_{\rm F}}.
\end{equation}
By expanding the equations (\ref{eq:ari}) and (\ref{eq:bri}) in the first order of 
$y$ and $|\omega_n|$,\cite{Nagai}  
we obtain the approximate solution as 
\begin{equation}
g(\rv,\kv_{\rm F}; {\rm i} \omega_n) \sim - \frac{ |\Vec{v}_{\rm F}| |\lambda(\kv_{\rm F})|e^{-2 \lambda F(x) }}{2 \xi_0 [{\rm i} E_{\rm ani} + \omega_n]},
\end{equation}
where
\begin{eqnarray}
E_{\rm ani}(\kv_{\rm F}) &\sim& \frac{y(\kv_{\rm F})}{\xi_0 h(\kv_{\rm F})} \Delta_{ \infty} \lambda^2(\kv_{\rm F}), \\
F(x) &=& \frac{1}{v_{\rm F \perp}h(\kv_{\rm F})} \int_0^{|x|} dx' f(x'),
\end{eqnarray}
with $\xi_0 = v_{\rm F}/(\pi \Delta_{\rm \infty})$.
Here, $f(x)$ describes the spatial variation of pair-potential and $f(0) = 0$, $\lim_{x \rightarrow 0} f(x) = \Delta_{\infty}$
 and $\lambda(\kv_{\rm F})$ describes the variation of pair-potential in momentum space and ${\rm MAX}(\lambda) = 1$.
The local density of states with the anisotropic Fermi surface is given by 
\begin{eqnarray}
\nu(\rv,\epsilon) 
                &\sim&  \int \frac{d S_{\rm F} v_{\rm F \perp} |\lambda| e^{- 2 \lambda F(x)}}{4 \pi^2 \xi_0 |\Vec{v}_{\rm F}|} 
                	 \delta(\epsilon - E_{\rm ani}) \label{eq:nua}.
\end{eqnarray}
To consider the smearing effects, we approximate Eq.~(\ref{eq:nua}) as 
\begin{equation}
\nu(\rv,\epsilon) \sim  \int \frac{d S_{\rm F}}{4 \pi^2 \xi_0 |\Vec{v}_{\rm F}|} 
        	\frac{v_{\rm F \perp} |\lambda| \delta e^{- 2 \lambda F(x)}}{(\epsilon - E_{\rm ani})^2 + \delta^2},
\end{equation}
with the smearing factor $\delta$.
Therefore, determining the distribution of the Fermi velocity $h(\kv_{\rm F})$ from the band structure and 
the distribution of the pair-potential $\lambda(\kv_{\rm F})$ from the gap structure, 
one can obtain the LDOS pattern around a vortex easily.
\section{In the case of ${\bf YNi_2B_2C}$}
\subsection{Local Density of States around a Vortex core}
\begin{figure}
\includegraphics[width = 8cm]{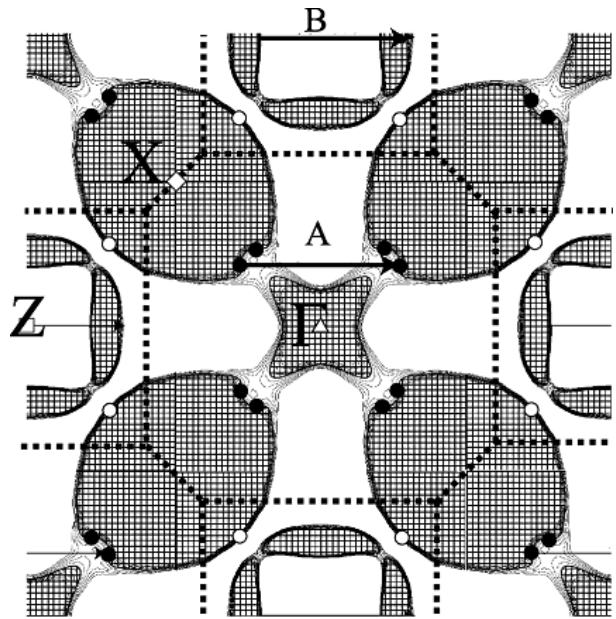}
\caption{\label{fig:gaps} The cross section  including the $\Gamma$ point of FS with the bct Brillouin zone for the 17th band.
A solid circle $\bullet$ and an open circle $\circ$ indicate the positions of point-node and the local minima, respectively.
An open triangle $\triangle$, box $\Box$ and diamond $\Diamond$ indicate the $\Gamma$ point, $Z$ point, $X$ point, respectively.}
\end{figure}

We calculate the LDOS pattern around a vortex with using the band structure calculated by 
Yamauchi {\it et al}.\cite{Yamauchi}
We consider the FS from 17th band and neglect the FSs from 18th and 19th bands 
since the values of the density of states for 17th, 18th, 19th band at Fermi energy are 
48.64, 7.88 and 0.38 states/Ry, respectively, so that 
the electrons of 17th band contribute dominantly to the superconductivity (See, Fig.~4 in Ref.~\onlinecite{Yamauchi}). 
The FS has two kinds of particular points on the planes $k_z  =0$ and $k_z = 0.5$.
We call the vectors at these points ^^ vector A' and ^^ vector B', respectively (See, Fig.~\ref{fig:gaps}).
The ^^ vector A' is the nesting vector and 
the ^^ vector B' is the vector which connects the pair with the antiferromagnetic fluctuations.
The Fermi velocities at these ^^ particular points' are in the $\langle 100 \rangle$ direction.
Maki {\it et al.} suggested that the pair-potential at these particular points given above is strongly suppressed 
because of an instability in the particle-hole channel which strongly depresses the effective potential 
for Cooper pairing.\cite{MakiThal}
We also investigate the positional relation between these particular points and point-nodes in momentum space 
to test Maki's scenario.

We assume that point-nodes are around the points which have the strong antiferromagnetic fluctuations.
We calculate the LDOS patterns assuming various gap structures where positions of point-nodes are different.
Comparing with the STM experiments, we obtain the most suitable gap structure as shown in Fig.~\ref{fig:gaps}.
In Fig.~\ref{fig:gaps},  solid circles $\bullet$ and open circles $\circ$ indicate the positions of point-node
 and the local minima, respectively.
It should be noted that point-nodes are only at the ^^ vector A' 
and the gap functions do not have point-node at the ^^ vector B'.
If the gap functions have point-node at the ^^ vector B', the LDOS patterns are not consistent with the STM experiments.

In Figs.~\ref{fig:LDOSs}(d)-(f), we show the calculated LDOS with this most suitable gap structure for several bias energies $\epsilon$.
It is seen from Fig.~\ref{fig:LDOSs}(d) that the fourfold star centered at a vortex core extends toward $\langle 100 \rangle$
 for $\epsilon/\Delta_{\infty} = 0$.
As shown in Figs.~\ref{fig:LDOSs}(e) and (f), 
the peak of the LDOS splits into four peaks toward $\langle$ 110 $\rangle$ with increasing energy.
These LDOS patterns in Figs.~\ref{fig:LDOSs}(d)-(f) coincide with the observation in Figs.~\ref{fig:LDOSs}(a)-(c), respectively.
If we assume the gap structure which does not have local minima at $\circ$, 
the calculated LDOS patterns are not consistent with the STM/STS experiments, since 
the peak of this calculated LDOS splits into four peaks toward $\langle$  100 $\rangle$ with increasing energy.
If we assume the isotropic gap structure with this FS, 
the calculated LDOS patterns are almost circle.
Our analytical theory in the previous paper \cite{Nagai} shows that 
the LDOS around a vortex core consists of the contribution of the quasiparticles with momentum where the pair-potential is large on the FS.
In other words, we can obtain the information of anti-node direction on FS from STM/STS experiments.
From our calculations of LDOS, we can show that 
the gap amplitude at $\bullet$ and $\circ$ are smaller than the half of the maximum gap.
Therefore, we need comparisons with FAD heat capacity and AV thermal transport experiments, from which we can 
obtain the information of gap-nodes.


\subsection{Angular Dependence of the Density of States in the Weak Magnetic Fields}
Comparing with the FAD heat capacity and the AV thermal conductivity experiments, 
we calculate the density of states with the Doppler-shift method.\cite{Volovik} 
Without impurity scattering (superclean limit), the quasiclassical Green function is written as 
\begin{equation}
g(\rv,\hat{\pv}; {\rm i} \omega_n) = \frac{\omega_n + {\rm i} \Vec{v}_{\rm F} \cdot \Vec{v}_{\rm s} }
	{\sqrt{(\omega_n + {\rm i} \Vec{v}_{\rm F} \cdot \Vec{v}_{\rm s})^2+|\Delta|^2}
        }.
\end{equation}
Here, $\Vec{v}_{\rm s}$ is the supercurrent velocity around a vortex.
Therefore, the density of states at zero-energy $\nu(\epsilon = 0)$ is written as 
\begin{equation}
\nu(\epsilon=0,\phi) = \Bigl{\langle} {\rm Re} \frac{
	d(\phi)
        				}
	{ \sqrt{
        	 d ^2(\phi) - |\Delta|^2 
                 }
        } \Bigl{\rangle}. \label{eq:doppler}
\end{equation}
Here, $d(\phi) =|\Vec{v}_{\rm F} \cdot \Vec{v}_{\rm s}(\phi)|$,
the bracket means averaging over both Fermi surface and unit cell of vortex lattice 
and $\phi$ is the angle between magnetic fields and the $a$ axis on the $a$-$b$ plane.
Since the core states do not contribute to the specific heat in low magnetic fields, 
we neglect the spatial variation of the magnitude of the pair-potential 
and consider the spatial variation of the phase of the pair-potential around a vortex.
Assuming $d \ll \Delta_{\infty}$, only the quasiparticles around nodes in momentum space  contribute to the density of states.
This assumption is appropriate in weak magnetic fields.

\begin{figure}
\includegraphics[width = 8cm]{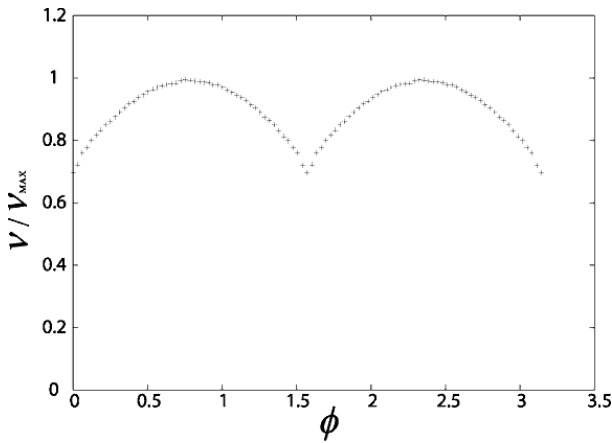}
\caption{\label{fig:gapdos} Angular dependence of specific heat $C_{\rm s} \propto \nu(\epsilon = 0,\phi)$
 with use of the band-structure 
of YNi$_2$B$_2$C. 
$\phi$ is the angle between magnetic field direction and the $a$ axis.}
\end{figure}

Figure~\ref{fig:gapdos} shows the angular dependence of specific heat $C_{\rm s} \propto \nu(\epsilon = 0,\phi)$ 
calculated with use of the band-structure of YNi$_2$B$_2$C and the gap-structure consistent with 
the STM experiments (See Fig.~\ref{fig:gaps}).
Here, the magnetic field is applied within the $a$-$b$ plane at an angle $\phi$ from the $a$ axis.
It should be noted that narrow cusps appear at $\phi = 0$ and $\phi = \pi/2$.
This angular variation is consistent with the FAD heat capacity 
and the AV thermal transport experiments as shown in Fig.~2 in Ref.~\onlinecite{Park} and Fig.~2 in Ref.~\onlinecite{Izawa}, respectively.
These angular variations can be explained by the fact 
that quasiparticles (QPs) around the gap-nodes in momentum space contribute to the DOS and 
QPs traveling parallel to the vortex do not 
contribute to the DOS (Note that QP with $\Vec{v}_{\rm F}$ parallel to the vortex line
gives no contribution in Eq.~(\ref{eq:doppler}) because $\Vec{v}_s$ is perpendicular to
the vortex line and $d=| {\bm v}_{\rm F} \cdot {\bm v}_s |=0$ for such a QP.). 
In other words, the narrow cusps appear at the direction of the Fermi velocity of QPs at the gap-nodes. 
In the case of YNi$_2$B$_2$C, the direction of the Fermi velocity at $\bullet$ is 
parallel to the $\langle$ 100 $\rangle$.
\subsection{Density of States without Magnetic Fields}
\begin{figure}
\includegraphics[width = 8cm]{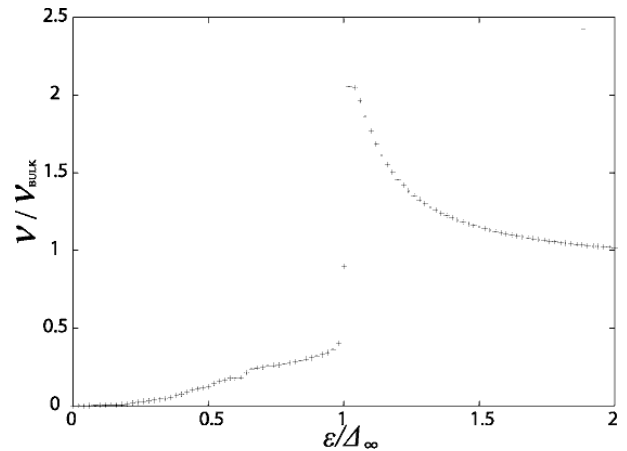}
\caption{\label{fig:DOS} Quasiparticle  density of states without magnetic fields. the smearing 
factor $\delta = 0.05 \Delta_{\infty}$.}
\end{figure}

Comparing with the result of the STM/STS experiment in the absence of magnetic fields,\cite{Nakai} 
we calculate the density of states in the absence of magnetic field in the bulk region.
In Fig.~\ref{fig:DOS}, we show the DOS 
with use of the band-structure and the gap-structure.
Here, we assume the smearing parameter $\delta = 0.05 \Delta_{\infty}$.
Any singularities like that at $E = 0.5\Delta_{\infty} $in the case of $s$+$g$ wave superconductivity\cite{MakiThal} 
do not occur in our gap-structure.
Our result is consistent with the density of states by the STM/STS experiment shown in Fig.~4(c) in Ref.\onlinecite{Nakai}.

\subsection{Angular Dependence of the LDOS patterns}
\begin{figure}
\includegraphics[width = 8cm]{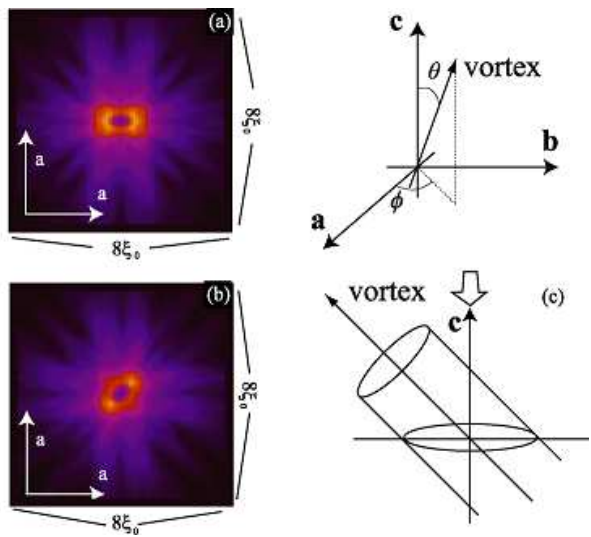}
\caption{\label{fig:per}(Color Online)  Distributions of the local density of states for the vortex tilting 
from the $a$ axis by angle $\theta$ 
and the $c$ axis by angle $\phi$.(a) $(\phi,\theta) = (0,\pi/4)$, (b) $(\phi, \theta) = (\pi/4,\pi/4)$. 
The smearing factor is $\delta = 0.05 \Delta_{\infty}$ and the energy is $\epsilon = 0.3 \Delta_{\infty}$.
These LDOS patterns will be observed by the STM/STS experiments from the $c$ axis. 
}
\end{figure}

One can obtain the three-dimensional information of the pairing symmetry from STM/STS experiments
under the magnetic field with various directions.
We have shown the LDOS patterns for the vortex parallel to the $b$ axis about $s$+$g$ wave superconductor in Fig.~14 in Ref.~\onlinecite{Nagai}.
We calculate the LDOS patterns on the $a$-$b$ plane with 
various directions of magnetic fields, 
since 
the LDOS will be observed by the STM/STS experiments on 
the clean surface perpendicular to the crystal $c$ axis.
As shown in Fig.~\ref{fig:per}, these LDOS patterns are different qualitatively.
The four peaks in the LDOS patterns for the vortex parallel to the $c$ axis disappear
in the LDOS patterns 
for the vortex tilting from the $c$ axis by angle 
$\theta = \pi/4$ within the $a$-$c$ plane as shown in Fig.~\ref{fig:per}(a).
Only the two peaks remain in the LDOS pattern for the vortex tilting from the $a$ axis by angle $\phi = \pi/4$ 
and tilting from the $c$ axis by angle $\theta = \pi/4$ as shown in Fig.~\ref{fig:per}(b).

\section{Discussions}
We considered the 17th band and neglect the 18th and 19th bands.
By the angular resolved photo emission experiment,\cite{Baba} the isotropic superconducting gap 
exists on the FS from 18th band 
and the FS from 19th band is not observed.
Since the LDOS pattern 
from the isotropic superconducting gap is almost isotropic, 
the LDOS from 18th band does not contribute to the four peak structure of the STM/STS experiment.
Comparing Fig.~\ref{fig:LDOSs}(c) with Fig.~\ref{fig:LDOSs}(f), the LDOS at the center by the theoretical calculation is quite smaller 
than that at a vortex center by the STM/STS experiments.
Considering the FS from 18th band, it seems that this LDOS at a vortex center becomes large.
Therefore, the calculations with use of the FSs from 17th and 18th bands are needed for quantitative comparison with STM/STS experiment.

The results of STM/STS experiments around a vortex can not be explained by the ^^ two-gap model' since 
the strong anisotropy of the gap-structure makes the four peak structure of this experiment. 
Two isotropic gaps make two isotropic LDOS patterns even if the highly anisotropic FS is considered.
The LDOS patterns with the model of point-nodes and that with the two-gap model are qualitatively different in the zero energy region 
as shown in Fig.~\ref{fig:LDOSs}(d).

The previous analyses of the FAD heat capacity and the AV thermal transport experiments are insufficient 
since the isotropic FS is assumed, so that these analyses lead to the wrong conclusion that 
the gap function has point-nodes along the $a$ axis and $b$ axis. 
The information about nodes we obtain is the directions of the Fermi velocity at the nodes on the FS, 
since the Doppler-shift method is based on the fact that the energy has the term 
proportional to $\Vec{v_{\rm F}} \cdot \Vec{v_{\rm s}}$ as shown in Eq.~\ref{eq:doppler}.
Therefore, these experiments suggest that the Fermi velocity at the nodes on the FS
 is parallel to the $a$ axis (the direction of nesting vectors). 
In the case of the anisotropic FS, the directions of the point-nodes are not found only from these experiments.
In the only case of the isotropic FS, the directions of the point-nodes are parallel to the Fermi velocity at the nodes.

We have shown that the point-nodes in momentum space are at the ^^ vector A'. 
These results suggest that the strong antiferromagnetic fluctuations suppress the superconducting order parameter 
at these points.\cite{MakiThal,Kontani}
The reason why the point-nodes do not exist at the ^^ the vector B' is not clear yet. 
This is a future problem.

\section{Conclusion}
In conclusion, we calculated the LDOS patterns around a vortex core, the angular dependence of the DOS in weak magnetic field, 
the DOS in zero fields in the bulk region and the LDOS patterns for the vortex tilting from the crystal $c$ axis 
with use of the FS from 17th band obtained by the band-calculations, assuming 
various gap structures where positions of point-nodes are different. 
Comparing our theoretical calculations with  the results of the STM/STS, the FAD heat capacity and the AV thermal transport experiments,
we determined the gap-structure which is consistent with these experiments.
The point-nodes are at ^^ the vector A', which is along $\langle 110 \rangle$ as shown in Fig.~\ref{fig:gaps}.
We also showed that the previous analyses of the FAD heat capacity and the AV thermal transport experiments are insufficient
since the isotropic FS was assumed.
Considering the anisotropic FS, we showed the most suitable gap structure. 
This gap structure is consistent with the results of the FAD heat capacity, 
the AV thermal transport, the density of states without magnetic fields by the STM
and the local density of states around a vortex core by the STM.
We hope that our results will be tested by the STM/STS experiments with the rotation of the magnetic fields.

\section*{Acknowledgment}
We thank M. Udagawa, S. Kaneko, N. Nishida, T. Shibauchi and Y. Matsuda for helpful discussions.
We also thank T. Baba for showing the latest experimental data.
This work is supported by a Grant-in-Aid for Scientific Research (C)(2) No. 17540314 from the Japan Society for the Promotion of Science.

\end{document}